\documentclass[12pt]{article}
\usepackage{epsf}
\def\be{\begin{equation}}
\def\ee{\end{equation}}
\def\bea{\begin{eqnarray}}
\def\eea{\end{eqnarray}}
\advance\textheight by \topskip
\def\la{\mathrel{\mathpalette\fun <}}
\def\ga{\mathrel{\mathpalette\fun >}}
\def\fun#1#2{\lower3.6pt\vbox{\baselineskip0pt\lineskip.9pt
  \ialign{$\mathsurround=0pt#1\hfil##\hfil$\crcr#2\crcr\sim\crcr}}}

\begin{document}


\begin{center}

{\bf \LARGE Inflation, Quantum Cosmology\\

\vskip 0.5cm

 and the Anthropic Principle\footnote{To appear in ``Science and Ultimate Reality: From Quantum to Cosmos", honoring John Wheeler's 90th birthday. J. D. Barrow, P.C.W. Davies, \& C.L. Harper eds. Cambridge University Press (2003)} } 

\

\

{\bf  Andrei Linde}

\vskip 0.2cm

 Department of Physics, Stanford University, Stanford,
CA 94305, USA 

\end{center}

\

\tableofcontents{} 

\newpage 

\parskip 5.5pt

\section{Introduction}
 
One of the main desires of physicists is to construct a theory
that unambiguously predicts the observed values for all
parameters of all elementary particles.  It is very tempting to
believe that the correct theory describing our world should be
both beautiful and unique. 

However, most of the parameters of elementary
particles look more like a collection of random numbers than a
unique manifestation of some hidden harmony of Nature. For example, the mass of the electron is 3 orders of magnitude smaller than the mass of the proton, which is 2 orders of magnitude smaller than the mass of the W-boson, which is 17 orders of magnitude smaller than the Planck mass $M_p$.
Meanwhile, it was pointed out long ago that a minor change (by a
factor of two or three) in the mass of the electron, the
fine-structure constant $\alpha_e$, the strong-interaction
constant $\alpha_s$, or the gravitational constant $G= M_p^{-2}$ would lead to
a universe in which life as we know it could never have arisen.
Adding or subtracting even a single spatial dimension  of the same type as the usual three dimensions would make 
planetary systems impossible. Indeed, in space-time with
dimensionality $d > 4$, gravitational forces between distant
bodies fall off faster than $r^{-2}$, and in
space-time with $d<4$, the general theory of relativity tells us
that such forces are absent altogether.
This rules out the existence of stable planetary systems for $d\not = 4$.
Furthermore, in order for life as we know it to exist, it is
necessary that the universe be sufficiently large, flat,
homogeneous, and isotropic.  These facts, as well as a number of
other observations, lie at the foundation of the
so-called anthropic principle (Barrow and Tipler, 1986; Rozental, 1988; Rees, 2000)
According to this principle, we observe the universe to be as it
is because only in such a universe could observers like ourselves
exist.  

Until very recently, many scientists  were ashamed of using the anthropic principle in their research. A typical attitude was expressed in the book ``The Early Universe'' by Kolb and Turner: ``It is unclear to one of the authors how a concept as lame as the ``anthropic idea'' was ever elevated to the status of a principle''  (Kolb, 1990).

This critical attitude is quite healthy. It is much better to find a simple physical resolution of the problem rather that speculate that we can live only in the universes where the problem does not exist. There is always a risk that the anthropic principle does not cure the problem, but acts  like a painkiller.

On the other hand, this principle can help us to understand that some of the most complicated and fundamental problems may become nearly trivial if one looks at them from a different perspective. Instead of denying the anthropic principle or uncritically embracing it, one should take a more patient approach and check whether it is really helpful or not  in each particular case.

There are two main versions of this principle: the weak anthropic principle and the strong one. The weak anthropic principle simply says that if the universe consists of different parts  with different properties, we will live only in those parts where our life is possible. This could seem rather trivial, but one may wonder whether these different parts of the universe are really available. 
If it is not so, any  discussion  of altering the mass of the electron, the fine
structure constant, and so forth is perfectly meaningless.

The strong anthropic principle says that the universe must be created in such a way as to make our existence possible. At first glance, this principle  must be
faulty, because mankind, having appeared $10^{10}$ years
after the basic features of our universe were laid down, could in
no way influence either the structure of the universe or the
properties of the elementary particles within it.  

Scientists often associated the anthropic principle  with the idea that the universe was created many times until the final success. It was not clear who did it and why was it necessary to make the universe  suitable for our existence. Moreover, it would be much simpler to create proper conditions for our existence in a small vicinity of a solar system rather than in the whole universe. Why would one need to work so hard?

Fortunately, most of  the  problems associated with the anthropic principle were resolved (Linde, 1983a,1984b,1986a)
soon after the invention of inflationary cosmology. Therefore we will  remember  here the basic principles of inflationary theory.

\section{Chaotic inflation}
Inflationary theory was formulated in many different ways, starting with the  models based on quantum gravity (Starobinsky, 1980) and on the theory of high temperature phase transitions with supercooling and exponential expansion in the false vacuum state (Guth, 1981; Linde, 1982a; Albrecht and Steinhardt, 1982). However, with the introduction of the chaotic inflation scenario (Linde,  1983b) it was realized that the basic principles of inflation actually are very simple, and no thermal equilibrium, supercooling, and expansion in the false vacuum is required.

To explain the main idea of chaotic inflation, let us consider  the
simplest model of a scalar field $\phi$ with a mass $m$ and with the
potential energy density $V(\phi)  = {m^2\over 2} \phi^2$,  see  Fig. 1.
Since this function has a minimum at $\phi = 0$,  one may expect that the
scalar field $\phi$ should oscillate near this minimum. This is indeed
the case if the universe does not expand. However, one can show that in a
rapidly expanding universe  the scalar field moves down very slowly, as a
ball in a viscous liquid, viscosity being proportional to the speed of
expansion.

There are  two equations which describe evolution of a homogeneous scalar
field
 in our model, the field equation
\begin{equation}\label{1}
\ddot\phi + 3H\dot\phi = -m^2\phi \ ,
\end{equation}
and the Einstein equation
\begin{equation}\label{2}
H^2 +{k\over a^2} ={8\pi \over 3M_p^2}\, \left(  {1\over 2}\dot \phi^2+V(\phi) \right) \ .
\end{equation}
Here $H = \dot a/a $ is the Hubble parameter in the universe with a scale
factor $a(t)$ (the size of the universe), $k = -1, 0, 1$ for an open, flat or closed universe
respectively, $M_p$ is the Planck mass, $M_p^{-2} = G$, where $G$ is the gravitational constant. The first equation becomes similar to the
 equation of motion for a harmonic oscillator, where instead of $x(t)$ we have
$\phi(t)$. The term  $3H\dot\phi$ is similar to the term describing friction in the equation for a harmonic oscillator.

\begin{figure}
\centering\leavevmode\epsfysize=8 cm \epsfbox{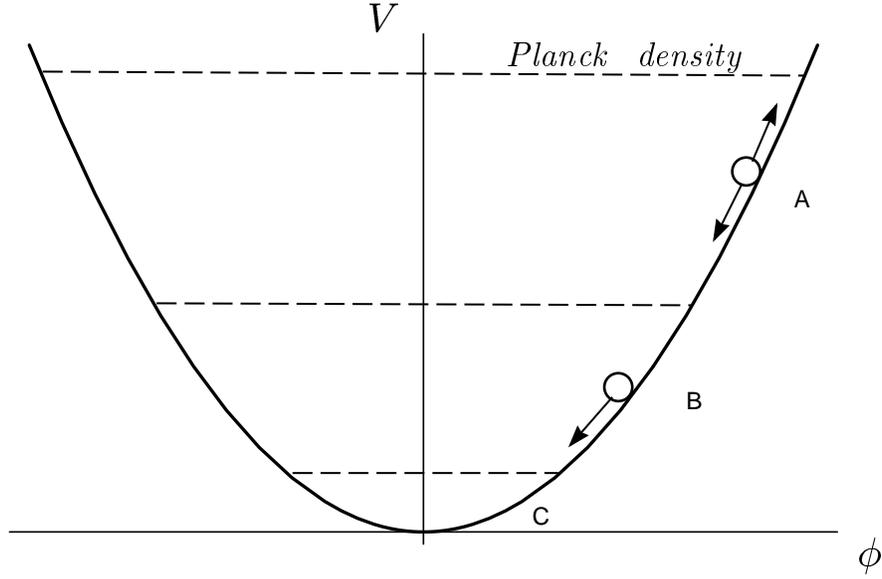}

\

\

\caption{Motion of the scalar field in the theory with $V(\phi) =
{m^2\over 2} \phi^2$. Several different regimes are possible,  depending
on the value of the field $\phi$. If the potential energy density of the
field is greater than the Planck density $\rho \sim M_p^4 \sim 10^{94}$ g/cm$^3$,
quantum fluctuations of space-time are so strong that one cannot describe
it in usual terms. Such a state   is called space-time foam. At a
somewhat smaller energy density  (region A: $m M_p^3 < V(\phi) < M_p^4$)
quantum fluctuations of space-time are small, but quantum fluctuations of
the scalar field $\phi$ may be large. Jumps of the scalar field due to
quantum fluctuations lead to a process of eternal self-reproduction of
inflationary universe which we are going to discuss later. At even
smaller values of $V(\phi)$ (region B: $m^2  M_p^2 < V(\phi) < m M_p^3$ )
fluctuations of the field $\phi$ are small; it slowly moves down as a
ball in a viscous liquid. Inflation occurs both in the region A and
region B. Finally, near the minimum of $V(\phi)$ (region C) the scalar
field rapidly oscillates, creates pairs of elementary particles, and the
universe becomes hot.} \label{fig:Fig1}
\end{figure}

 If   the scalar field $\phi$  initially was large,   the Hubble parameter $H$
was large too, according to the second equation. This means that the
friction term  was very large, and therefore    the
scalar field was moving   very slowly, as a ball in a viscous liquid.
Therefore at this stage the energy density of the scalar field, unlike
the  density of ordinary matter,   remained almost constant, and
expansion of the universe continued with a much greater speed than in the
old cosmological theory. Due to the rapid growth of the scale of the
universe and a slow motion of the field $\phi$, soon after the beginning
of this regime one has $\ddot\phi \ll 3H\dot\phi$, $H^2 \gg {k\over
a^2}$, $ \dot \phi^2\ll m^2\phi^2$, so  the system of equations can be
simplified:
\begin{equation}\label{E04}
 3{\dot a \over a}\dot\phi = -{m^2\phi} \ ,   
\end{equation} 

\begin{equation}\label{E04aaa}
H  ={\dot a \over a}   ={2
m\phi\over M_p}\, \sqrt { \pi  \over 3} \   .
\end{equation}
The last equation shows that the size of the universe $a(t)$ in this regime
grows approximately as $e^{Ht}$, where $H = {2 m\phi\over M_p}\, \sqrt {
\pi  \over 3}\,  $.

This stage of   exponentially rapid expansion of the universe is called
inflation.   In  realistic versions of inflationary theory its duration
could be as short as $10^{-35}$ seconds. When the   field $\phi$ becomes
sufficiently small,  viscosity becomes small,  inflation ends, and  the
scalar field $\phi$ begins to   oscillate near the minimum of $V(\phi)$.
As any rapidly oscillating classical field, it looses its energy by
creating pairs of elementary particles. These particles interact with
each other and come to a state of thermal equilibrium with some
temperature $T$. From this time on, the corresponding part of the
universe can be described by the standard hot universe theory.

The main difference between inflationary theory and the old cosmology
becomes clear when one calculates the size of a typical inflationary
domain at the end of inflation. Investigation of this issue shows
that even if  the initial size of   inflationary universe  was as small
as the Plank size $l_P \sim 10^{-33}$ cm, after $10^{-35}$ seconds of
inflation   the universe acquires a huge size of   $l \sim 10^{10^{12}}$
cm. This makes our universe  almost exactly flat and homogeneous on  large scale because all inhomogeneities were stretched by a factor of $10^{10^{12}}$.

This number is model-dependent, but in all realistic models the  size of
the universe after inflation appears to be many orders of magnitude
greater than the size of the part of the universe which we can see now,
$l \sim 10^{28}$ cm. This immediately solves most of the problems of the
old cosmological theory (Linde,  1990a).

Consider a universe which initially consisted of many domains
with chaotically distributed scalar field  $\phi$ (or if one considers
different universes with different values of the field). Those  domains where the scalar field was too small never inflated, so they do not contribute 
much to the total volume of the universe. The main
contribution to the total volume of the universe will be given by those
domains which originally contained large scalar field $\phi$. Inflation
of such domains creates huge homogeneous islands out of the initial chaos, each  homogeneous domain being much greater than the size
of the observable part of the universe. That is why I called this scenario `chaotic inflation'.

There is a big difference between this scenario and the old idea that the whole universe was created at the same moment of time (Big Bang), in a nearly uniform state with indefinitely large temperature.  In the new theory, the
condition of uniformity and thermal equilibrium is no longer required. Each part of the universe could have a singular beginning (see (Borde {\it et al},  2001) for a recent discussion of this issue). However,  in the context of chaotic inflation, this does not mean that the universe as a whole had a single beginning. Different parts of the universe could come to existence at different moments of time, and then grow up to the size much greater than the total size of the universe. The existence of initial singularity (or singularities) does not imply that the whole universe was created simultaneously in a single Big Bang explosion. In other words, we cannot tell anymore that the whole universe was born at some time $t=0$ before which it did not exist. This conclusion is valid for all versions of chaotic inflation, even if one does not take into account the process of self-reproduction of the universe discussed in Section \ref{eter}.

The possibility that our homogeneous part of the universe emerged from the chaotic state initial state has  important implications for the anthropic principle. Until now we have considered the simplest inflationary model with only
one scalar field.
Realistic models of elementary particles involve many other  
 scalar fields.  For example, according to the standard theory of electroweak interactions, masses of all elementary particles depend on the  value of the  Higgs scalar field $\varphi$ in our universe. This value is determined  by the position of the minimum of the effective potential ${V}(\varphi)$ for the field  $\varphi$. In the simplest models, the potential $V(\varphi)$ has only one minimum.  However, in general, the potential $V(\varphi)$ may have many different minima. For example, in the simplest supersymmetric theory unifying weak, strong and electromagnetic interactions, the effective potential has several different minima of equal depth with respect to the two scalar fields, $\Phi$ and $\varphi$. If the scalar fields $\Phi$ and $\varphi$ fall  to different minima in different parts of the universe (the process called spontaneous symmetry breaking), the masses of elementary particles and the laws describing their interactions  will be different in these parts. Each of these parts may become exponentially large because of inflation. In some of these parts, there will be no difference between weak, strong and electromagnetic interactions, and life of our type will be impossible there. Some other parts will be similar to the one where we live now (Linde, 1983c). 

 This means that even if we will be able to find the final theory of everything, we will be unable to  uniquely determine properties of
elementary particles in our universe; the universe may consist of different  exponentially large domains  where the properties of elementary particles may be different. This is an important step towards the justification of the anthropic principle. A further step can be made if one takes into account quantum fluctuations produced during inflation.

\section{Inflationary quantum fluctuations}

According to quantum field theory, empty space is not entirely
empty. It is filled with quantum fluctuations of all types of
physical fields.   The wavelengths of all quantum
fluctuations of the scalar field $\phi$ grow exponentially during
inflation. When the wavelength of any particular fluctuation becomes
greater than $H^{-1}$, this fluctuation stops oscillating, and its
amplitude freezes at some nonzero value $\delta\phi (x)$ because of the
large friction term $3H\dot{\phi}$ in the equation of motion of the field
$\phi$. The amplitude of this fluctuation then remains almost unchanged
for a very long time, whereas its wavelength grows exponentially.
Therefore, the appearance of such a frozen fluctuation is equivalent to
the appearance of a classical field $\delta\phi (x)$ produced from quantum fluctuations.

Because the vacuum contains fluctuations of all wavelengths, inflation
leads to the continuous creation of  new perturbations of the classical
field with wavelengths greater than $H^{-1}$. An average amplitude of perturbations generated during a time interval $H^{-1}$
(in which the universe expands by a factor of e) is given by $
|\delta\phi(x)| \approx \frac{H}{2\pi}$ (Vilenkin and Ford, 1982; Linde, 1982c).

These quantum fluctuations are responsible for galaxy formation (Mukhanov and Chibisov, 1981; Hawking, 1982; Starobinsky, 1982; Guth and Pi, 1982; Bardeen {\it et al}, 1983). But if the Hubble constant during inflation is sufficiently large,  quantum fluctuations of the scalar fields 
may lead not only to formation of galaxies, but also to the division of the universe into exponentially large domains with different properties.

As an example, consider again  the simplest supersymmetric theory unifying weak, strong and electromagnetic interactions. Different minima of the effective potential in this model are separated from each other by the distance $\sim 10^{-3} M_p$. The amplitude of quantum fluctuations of the fields $\phi$,  $\Phi$ and $\varphi$ in the beginning of chaotic inflation can be  as large as $10^{-1} M_p$.  This means that at the early stages of inflation the fields $\Phi$ and $\varphi$ could easily jump from one minimum of the potential to another.  Therefore even if initially these fields occupied the same minimum all over the universe, after the stage of chaotic inflation the universe becomes divided into many exponentially large domains corresponding to all possible minima of the effective potential (Linde, 1983c, 1984b). 

\section{Eternal chaotic inflation}\label{eter}

The process of the division of the universe into different parts becomes even easier if one takes into account the process of self-reproduction of inflationary domains. The basic mechanism can be understood as follows. If quantum fluctuations are sufficiently large, they may locally increase
the value of the potential energy of the scalar field in some parts of the universe. The probability of quantum jumps leading to a local increase of the energy density can be very small, but the  regions where it happens start expanding much faster than their parent domains, and quantum fluctuations inside them lead to production of new inflationary domains which expand even faster. This surprising behavior leads to the process of self-reproduction of the universe.  

This process is possible in the new inflation scenario (Steinhardt, 1982; Linde, 1982a; Vilenkin, 1983). However, even though the possibility to use this result for the justification of the anthropic principle was mentioned in (Linde, 1982a), this observation did not attract much attention because the  amplitude of the fluctuations in new inflation typically is  smaller than $10^{-6} M_p$. This is too small to probe most of the vacuum states available in the theory. As a result,  the existence of the self-reproduction regime in the new inflation scenario was basically forgotten; for many years this effect was not studied or used in any way even by those who have found it.

The situation changed dramatically when it was found that the self-reproduction of the universe occurs not only in new inflation but also in the   chaotic inflation scenario (Linde,  1986a).  In order to understand this effect,  let us consider an
inflationary domain of initial radius $H^{-1}$
containing a sufficiently homogeneous field with initial
value $\phi \gg M_p$.
Equations (\ref{E04}), (\ref{E04aaa}) tell  us
that during a typical time interval $\Delta t=H^{-1}$ the field
inside this domain will be reduced by
$\Delta\phi = \frac{M_p^2}{4\pi\phi}$. Comparing this expression with
the amplitude of quantum fluctuations $\delta\phi \sim {H\over 2\pi}=  {m\phi\over\sqrt {3\pi} M_p}$,  one can easily see
that  for $\phi\gg \phi^*\sim {M_p\over 2} \sqrt{M_p\over m}$, one has $|\delta\phi|\gg |\Delta\phi|$, i.e. the motion of the field $\phi$ due to its quantum fluctuations
is much more rapid than its classical motion. 

During the typical time $H^{-1}$ the size of the domain of initial size $H^{-1}$ containing the field $\phi \gg \phi^*$ grows $e$ times, its volume increases $e^3 \sim 20$ times, and almost in a half of this new volume the field $\phi$ jumps up instead of falling down. Thus the total volume of inflationary domains containing the field $\phi\gg \phi^*$ grows approximately 10 times.  During the next time interval $H^{-1}$ this process continues; the universe enters an eternal process of self-reproduction.  I called this process `eternal inflation.' 

In this scenario the scalar field may wander for an indefinitely long time at the density approaching the Planck density. This induces quantum fluctuations of all other scalar field, which may jump from one minimum of the potential energy to another for an unlimited time. The amplitude of these quantum fluctuations can be extremely large, $\delta\varphi\sim \delta\Phi \sim 10^{-1} M_p$. As a result, quantum fluctuations generated during eternal chaotic inflation can penetrate through any barriers, even if they have Planckian height, and the universe after inflation becomes divided into indefinitely large number of exponentially large domains containing matter in all possible states corresponding to all possible mechanisms of spontaneous symmetry breaking, i.e. to the different laws of the law-energy physics (Linde,  1986a;  Linde {\it et al}, 1994).

A rich  spectrum of
possibilities may appear  during inflation in Kaluza-Klein and superstring
theories, where an exponentially large variety of vacuum states and 
ways of compactification is available for the original
10- or 11-dimensional space. The type of compactification determines
coupling constants, vacuum energy, symmetry breaking, and
finally, the effective dimensionality of the space we live in. As it was shown in (Linde and Zelnikov, 1988), chaotic inflation at a nearly Planckian density  may lead to a local change of the number of compactified dimensions; the universe becomes divided into exponentially large parts with different dimensionality. 

Sometimes one may have a continuous spectrum of various possibilities. For example, in the context of the Brans-Dicke theory, the effective gravitational constant is a function of the Brans-Dicke field, which also experienced fluctuations during inflation. As a result, the universe after inflation becomes divided into exponentially large parts with {\it all} possible values of the gravitational constant $G$ and the amplitude of density perturbations ${\delta\rho\over \rho}$ (Linde, 1990b; Garcia-Bellido {\it et al} 1994). Inflation may divide our universe into exponentially large domains with continuously varying baryon to photon ratio ${n_B\over n_\gamma}$ (Linde, 1985) and with galaxies having vastly different properties (Linde, 1987b). Inflation may also continuously change the effective value of the vacuum energy (the cosmological constant $\Lambda$), which is a pre-requisite for many attempts to find an anthropic solution of the cosmological constant problem (Linde,  1984b,1986b; Weinberg, 1987; Efstathiou, 1995; Vilenkin, 1995b; Martel {\it et al}, 1998; Garriga and Vilenkin, 2000,2001b,2002; Bludman and Roos, 2002; Kallosh and Linde, 2002). Under these circumstances, the most
diverse sets of  parameters of particle physics (masses, coupling constants, vacuum energy,
etc.) can appear after inflation.

To illustrate the possible consequences of such theories in the context of inflationary cosmology, we present here the results of computer
simulations of evolution of a system of two scalar fields during chaotic
inflation (Linde {\it et al}, 1994). The field $\phi$ is the inflaton field driving inflation; it is shown by the height of the distribution of the field $\phi(x,y)$ in a
two-dimensional slice of the universe. The field $\chi$ determines the
type of spontaneous symmetry breaking which may occur in the theory. We
paint the surface black if this field is in a state corresponding to one
of the two minima of its effective potential;  we paint it white if it is
in the second minimum corresponding to a different type of symmetry
breaking, and therefore to a different set of laws of low-energy physics.

In the beginning of the process the whole inflationary domain was black,
and the distribution of both fields was very homogeneous. Then the domain
became exponentially large and it became divided into exponentially large domains with different properties, see Fig. 2.  Each peak of the `mountains' corresponds
to a nearly Planckian density and can de interpreted as a beginning of a
new  Big Bang.  The laws of physics are rapidly changing there, but
they become fixed in the parts of the universe where the field $\phi$
becomes small. These parts correspond to valleys in Fig. 2. Thus quantum
fluctuations of the scalar fields divide the universe into exponentially
large domains with different laws of low-energy physics, and with
different values of energy density.

 \begin{figure} 
 \centering \leavevmode\epsfysize=14 cm  \epsfbox{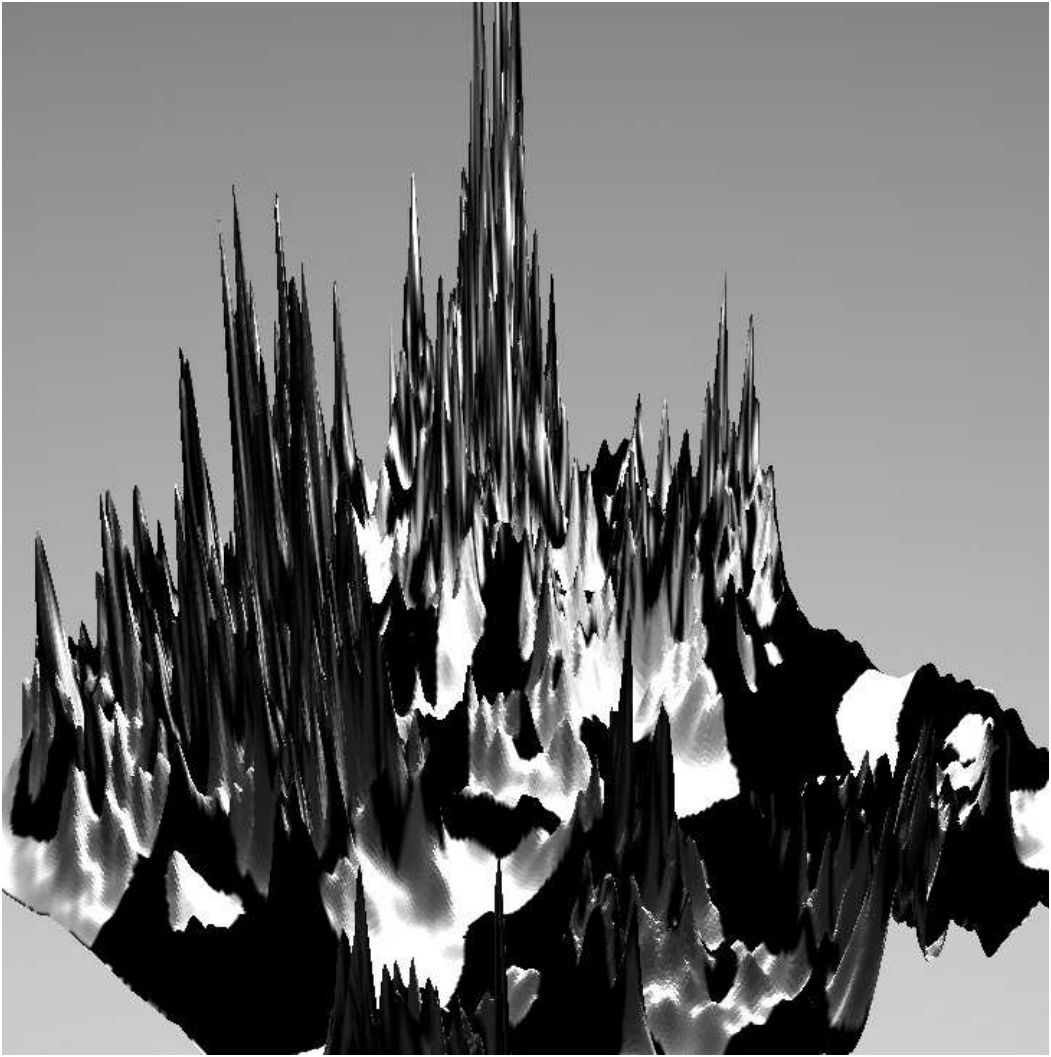}

\caption{A typical distribution of scalar fields  $\phi$ and $\chi$ during the process
of self-reproduction of the universe.   The height of the distribution
shows the value of the field $\phi$ which drives inflation. The surface
is painted black in those parts of the universe where the scalar field
$\chi$ is in the first minimum of its effective potential, and  white
where it is in the second minimum. Laws of low-energy physics are
different in the regions of different color. The peaks of the
`mountains' correspond  to places where quantum fluctuations bring the
scalar fields back to the Planck density. Each of such places in a
certain sense can be considered as a beginning of a new Big Bang. }
\label{fig:Fig0}
\end{figure}

    As a result of quantum jumps of the scalar fields during eternal inflation,   the universe  becomes divided into infinitely many exponentially large
domains with different laws of low-energy physics. Each of these domains is so large that for all practical purposes it can be considered a separate universe: Its inhabitants will live exponentially far away from its boundaries, so they will never know anything about the existence of other `universes' with different properties.

If this scenario is correct, then physics alone cannot provide a complete
explanation for all properties of our part of the universe.   The same
physical theory may yield large parts of the universe that have diverse
properties.  According to this scenario, we find ourselves inside a
four-dimensional domain with our kind of physical laws not because
domains with different dimensionality and with alternate properties are
impossible or improbable, but simply because our kind of life cannot
exist in other domains. 

This provides a simple justification of the weak anthropic principle  and removes the standard objections against it. One does not need anymore to  assume that some supernatural cause created our universe with the properties specifically fine-tuned to make our existence possible. Inflationary universe itself, without any external intervention, may  produce exponentially large domains with all possible laws of low-energy physics.  And  we should not be surprised that the conditions necessary for our existence appear on a very large scale rather than only in a small vicinity of the solar system.  If the proper conditions  are established near the solar system, inflation ensures that similar conditions appear everywhere within the observable part of the universe. 

The new possibilities that appear due to the self-reproduction of the universe may provide a basis for what I called `the Darwinian approach to cosmology' (Linde, 1987a; Vilenkin, 1995; Garcia-Bellido and Linde, 1995). Mutations of the laws of physics may lead to formation of the domains with the laws of physics that allow a greater speed of expansion of the universe; these domains will acquire greater volume and may host a greater number of observers. 

On the other hand, the total volume of domains of each type grows indefinitely large. This process looks like a peaceful coexistence and competition, and sometimes even like a fruitful collaboration, when the fastest growing domains produce many slower growing brothers. In this case a stationary regime is reached, and the speed of growth of the total volume of domains of each type becomes equally large for all of the domains (Linde {\it et al}, 1994).

\section{Baby universes}

As we have seen, inflation allows one to justify the weak anthropic principle by ensuring that all vacuum states and, consequently, all possible laws of elementary particle physics that are allowed by the basic theory are realized in some exponentially large and locally uniform parts of our universe.

Note, however, that here we are talking not about the choice among many different theories, but about the choice among many possible vacuum states, or phases, that are allowed by a given theory. This is similar to the possibility to find water in a gaseous, liquid or solid state. These states look very different (fish cannot live in ice), but their basic chemical composition is the same. Similarly, despite the fact that some of the theories may have extremely large number of vacuum states, our freedom of choice is still limited by the unique fundamental law that is supposed to remain the same in every corner of our universe.

Now it's time to make the next  step and ask whether the basic theory was in fact  fixed from the very beginning and could not change? 
A very interesting set of  ideas related to this question was developed in the end of the 80's. It was called the baby universe theory  (Coleman, 1988a,1988b; Banks, 1988; Giddings and Strominger, 1988,1989).  For a short time, this theory was immensely popular, but then it was almost completely forgotten. In our opinion, both extremes were due to the over-reaction with respect to the uncritical use of the   Euclidean approach to quantum cosmology. But if one distinguishes between this method and the rest of the theory, one can find something very interesting and instructive.

The main idea of the baby universe theory is that our universe can  split into disconnected pieces due to quantum gravity effects. Baby universes created from
the parent universe can carry from it an electron-positron pair,
or some other combinations of particles and fields, unless it is
forbidden by conservation laws. Such a process can occur in any place
in our universe. Many ways were suggested to describe such a
situation. The simplest one is to say that the existence of baby
universes leads to a modification of the effective Hamiltonian
density.
\begin{equation}\label{eq49}
{\cal H}(x) = {\cal H}_{0}(\phi(x)) + \sum{\cal H}_{i}[\phi(x)]
A_i \ .
\end{equation}
The Hamiltonian  (\ref{eq49}) describes the fields $\phi(x)$ on the parent
universe at distances much greater than the Planck scale. ${\cal H}_0$
is the part of the Hamiltonian which does not involve topological
fluctuations. ${\cal H}_{i}(\phi)$ are some local functions of the
fields $\phi$, and $A_i$ are combinations of creation and annihilation
operators for the baby universes. These operators do not depend on $x$
since the baby universes cannot carry away momentum. Coleman argued
(Coleman, 1988a,1988b) that the demand of locality, on the parent universe,
\begin{equation}\label{eq50}
[{\cal H}(x), {\cal H}(y)] = 0
\end{equation}
for spacelike separated $x$ and $y$, implies that the operators $A_i$
must all commute. Therefore, they can be simultaneously diagonalized
by the $``\alpha-states"$:
\begin{equation}\label{eq51}
A_i |{\alpha_i}\rangle = \alpha_i |{\alpha_i}\rangle \ .
\end{equation}
If the state of the baby universe is an eigenstate of the $A_i$,
then the net effect of the baby universes is to introduce infinite
number of undetermined parameters (the $\alpha_i$) into the
effective Hamiltonian (\ref{eq49}): one can just replace the operators $A_i$ by
their eigenvalues. If the universe initially is not in the $A_i$
eigenstate, then, nevertheless, after a series of measurements the
wave junction soon collapses to one of the $A_i$ eigenstates
(Coleman, 1988a,1988b; Giddings and Strominger, 1988,1989).

This gives rise to an extremely interesting possibility related to the
basic principles of physics. We were accustomed to believe that the
main purpose of physics is to discover the Lagrangian (or Hamiltonian)
of the theory that correctly describes our world. However, the question
arises: if our universe did not exist sometimes in a distant past,
in which sense could one speak about the existence of the laws of
Nature which govern the universe? We know, for example, that the laws of
our biological evolution are written in our genetic code. But where
were the laws of physics written at the time when there was no
universe (if there was such time)? The possible answer now is that the
final structure of the (effective) Hamiltonian becomes fixed only
after measurements are performed, which determine the values of
coupling constants in the state in which we live. Different effective Hamiltonians describe different laws of
physics in different (quantum) states of the universe, and by making
measurements we reduce the variety of all possible laws of physics to
those laws that are valid in the (classical) universe where we live.

We will not discuss this issue here any further, since it would require
a thorough discussion of the difference between the orthodox
(Copenhagen) and the many-world interpretation of quantum mechanics.
We would like to mention only that this theory  opens a new interesting
possibility to strengthen  the anthropic principle by allowing all fundamental constants to take different values  in different quantum states of the universe.

But if it is so interesting, why don't we hear about this theory anymore? In order to answer this question we must remember why it became so popular in the end of the 80's. The most interesting application of this theory was  the possible explanation of the vanishing of the cosmological constant (Coleman, 1988a,1988b).
The main idea is closely related to the previous suggestion by Hawking. According to (Hawking, 1984), the cosmological constant, like other
constants,  can take different values, and the probability to find ourselves
in the universe with the cosmological constant $\Lambda = V(\phi)$ is given by
\begin{equation}\label{eq52}
P(\Lambda) \sim \exp (-2S_{E}(\Lambda)) = \exp\frac{3 \pi  M^4_P}{\Lambda}\ ,
\end{equation}
where $S_E$ is the action in the  Euclidean version of  de Sitter space.
 However, Coleman pointed out that one should not only take into account one-universe Euclidean configurations. Rather one should sum over all configurations of babies and parents  connected by Euclidean wormholes. This finally gives
(Coleman, 1988a,1988b)
\begin{equation}\label{eq53}
P(\Lambda) \sim \exp \left(\exp \frac{3 \pi  M^4_P}{\Lambda}\right)   .
\end{equation}
Eqs. (\ref{eq52}) and  (\ref{eq53})       suggest  that it is most probable to live in a quantum
state of the
universe with $\Lambda = 0$. This would be a wonderful solution of the cosmological constant problem.

Unfortunately, the use of the Euclidean approach in this context was not well justified. The whole trick was based on the fact that Euclidean action $S_E$ has a wrong (negative) sign (Hartle and Hawking, 1983). Usually Euclidean methods work well for $S_E > 0 $ and become very problematic for $S_E<0$ (Linde, 1984a,1998; Vilenkin, 1984). After playing with this method for a while, most of the people became dissatisfied and abandoned it. Sometimes one can obtain sensible results by replacing $S_E$ by $|S_E|$ (Linde, 1984a; Vilenkin, 1984), but this would not yield any interesting results with respect to $\Lambda$ in the context of the baby universe theory.   Moreover, current observations suggest that the cosmological constant $\Lambda$ may be non-vanishing. As a result, the baby universe theory was nearly forgotten.  

From our point of view, however, the basic idea that the universe may exist in different quantum states corresponding to different laws of physics may be very productive. But this idea is still somewhat complicated because it pre-assumes that one can deal with the issues like that only at the level of the so-called third quantization (Coleman, 1988a,1988b; Banks, 1988; Giddings and Strominger, 1988,1989), with quantum field theory  applied not only to particles but also to the universes. This is a rather radical assumption. A somewhat different approach to quantum cosmology and variation of fundamental constants was suggested later in (Linde, 1990a; Vilenkin, 1995; Garcia-Bellido and Linde, 1995). Still it was usually emphasized that  these approaches are based on quantum cosmology, which is a rather complicated and controversial science.
Thus, it would be helpful to simplify these ideas a bit, and to present them in an alternative form that may allow further generalizations.

\section{From the universe to the multiverse}\label{multi}

Usually one  describes a physical theory by presenting its action. One may write, for example,  
\begin{eqnarray}\label{eq43}
S  = N\int d^4x \sqrt{g(x)}  
  \left(\frac{R(x)}{16\pi G} +L(\phi(x))\right),
\end{eqnarray}
where $N$ is a normalization constant, $\frac{R(x)}{2G} $ is the general relativity Lagrangian with $G = M_p^{-2}$, and  $L(\phi)$ is a Lagrangian for the usual matter fields. One obtains the Einstein equations by  variation of the action $S$ with respect to the metric $g^{\mu\nu}$, and one finds the equations of motion for the matter fields $\phi$ by variation of the action $S$ with respect to  $\phi$. 

Let us now do something very unusual and add  to our original action many other actions describing different fields $\phi_i$ with different Lagrangians $L_i$ living  in $k$ different universes of different dimensions $n_i$ with different metrics $g_i^{\mu\nu}$ and different gravitational constants $G_i$:
\begin{eqnarray}\label{eq43a}
S  &=& N\int d^4x \sqrt{g(x)}  
  \left(\frac{R(x)}{16\pi G}  +L(\phi(x))\right)\\ &+&\sum_{i=1}^{k} N_i\int d^{ {n_{i}}}x_i \sqrt{g_i(x_i)}  
  \left(\frac{R(x_i)}{16\pi G_i}  +L_i(\phi_i(x_i))\right). \nonumber
\end{eqnarray}
One may wonder whether this modification will affect our life in the universe described by the original action (\ref{eq43})? The answer is that it will have no impact whatsoever on the physical processes in our universe. Indeed, equations of motion for  $\phi$ and $g^{\mu\nu}$ will not change because the added parts do not depend on $\phi$ and $g^{\mu\nu}$, so their variation with respect to $\phi$ and $g^{\mu\nu}$ vanishes.

This implies that the extended action (\ref{eq43a}) describes all events in our universe  in the same way as the original action (\ref{eq43}). This is very encouraging. So let us continue our exercise and add to this action an {\it infinite} sum of {\it all} possible actions describing all possible versions of quantum field theory and M/string theory. If our original theory successfully described our universe, it will continue doing so even after all of these modifications.

But why would anybody want to add all of these extra terms if  they do not affect our universe? 

There are two related answers. First of all, one may simply reply: Why not?  
In some countries,  everything that is not explicitly allowed, 
 is forbidden. In  some other countries (and in science), 
 everything that is not explicitly forbidden, is  allowed. 
 We live in one of such countries, so why don't we use the freedom if it does not make us any harm?

But the second answer is more interesting. Now  we know that the theory (\ref{eq43a}) and all of its possible extensions are {\it exactly equivalent} to the theory (\ref{eq43}) with respect to the processes in our universe (assuming that it is described by (\ref{eq43})). So we can take a step back, look at all the different universes described by  Eq. (\ref{eq43a}), just  as we would look for our car among  many different cars in a parking lot, and ask: 

{\it As a matter of fact, which one of these universes is ours? Are we sure that it is the first one?}
 
From a purely theoretical point of view, the first  universe described by the theory (\ref{eq43}) is not any better  than any other  universe. However, we can live only in those universes that are compatible with the  existence of life as we know it. When we will search for our universe, first of all  we will look for those   Lagrangians $L_i(\phi_i(x_i))$  that  can describe elementary particles similar to the ones that we see around. Then we will specify  our search even further  by finding the Lagrangians describing particles with masses and coupling constants that are consistent with our existence.  Since we have {\it all} universes with {\it all } possible laws of physics described by our extended action, we will certainly find the universe where we can live. But that is exactly what we need to justify the validity of the strong anthropic principle.

Let us summarize our progress so far. Inflationary theory allows our universe to be divided into different parts with different laws of low-energy physics that are allowed by the unique fundamental theory. Most importantly, it makes each of such domains exponentially large, which is a necessary part of justification of the anthropic principle. The diversity of possible laws of physics can be very high, especially in the models of eternal chaotic inflation  where quantum fluctuations can have an extremely large amplitude, which makes the transition between all possible states particularly easy.

In addition to that, one can consider different universes with different laws of physics in each of them. This does not necessarily require introduction of quantum cosmology, many-world interpretation of quantum mechanics, and baby universe theory. It is sufficient to consider an extended action represented by a sum of all possible actions of all possible theories in all possible universes. One may call this structure a `multiverse.' This could sound like a very complicated and radical proposal, but in fact it is pretty trivial since each part of the infinite sum does not affect other parts. However, it establishes a firm formal background for the further development of the anthropic principle.

But the main reason why we are introducing this structure is not the anthropic principle. As we already mentioned, we need to know what emerged first at the moment of the universe formation: the universe or the law describing the universe. It is equally hard to understand how any law could exist prior to the universe formation, or how the universe could exist without a law. One may assume that there is only one possible law, and it exists in some unspecified way even prior to the emergence of the universe. However, this would be similar to having elections with only one name on the ballot. Perhaps a better possibility would be to consider all logically possible combinations of the universes, the laws describing them, and the observers populating these universes. Given the choice among different universes in this multiverse structure, we can proceed by eliminating the universes where our life would be impossible. This simple step is sufficient for understanding of  many features of our universe that otherwise would seem miraculous.

There are some additional steps that one may want to make. In our analysis we still assumed that any evolution must be described by some kind of action. Meanwhile there are some theories where equations of motion are known even though the action is unavailable. One may consider other models of evolution, based, e.g., on cellular automata. One can go even further, and consider all possible mathematical structures (Tegmark,  1998), or, following Wheeler, consider all logical possibilities and the concept of `it from bit'; see (Wheeler, 1990) and references therein.

But before doing so we would like to show that the concept of a multiverse may have interesting consequences going beyond the justification of the anthropic principle. In order to do it we must learn whether the different universes may interact with each other.

 \section{Double universe model and the cosmological constant problem}
Let us consider
the  double-universe model (Linde, 1988). This model describes
two universes, X and Y, with coordinates $x_\mu$ and
$y_\alpha$,
            respectively $(\mu, \alpha=0,1,\dots, 3)$ and with metrics
$g_{\mu\nu}(x)$ and $\bar{g}_{\alpha\beta}(y)$, containing fields
$\phi(x)$ and $\bar{\phi}(y)$ with the action of the
following unusual type:
\begin{eqnarray}\label{eq43bbb}
S  = N\int d^4xd^4y\sqrt{g(x)}\sqrt{\bar{g}(y)}\nonumber\\
 \times [\frac{M^2_p}{16\pi} R(x)+L(\phi(x))-\frac{M^2_p}{16\pi}
R(y) - L(\bar{\phi}(y))] \ .
\end{eqnarray}
Here $N$ is some normalization constant. This action is
invariant under general coordinate transformations in each
of the universes separately. A novel symmetry of the action
is the symmetry under the transformation $\phi(x)\rightarrow
\bar{\phi}(x), g_{\mu\nu}(x) \to
                     \bar{g}_{\alpha\beta}(x)$ and under the
subsequent change of the overall sign, $S\rightarrow -S$.
We call this the antipodal symmetry, since it relates to
each other the states with positive and negative energies.

An immediate consequence of this symmetry is the invariance
under the change of the values of the effective potentials
$V(\phi) \rightarrow V(\phi) + c,
               V(\bar{\phi}) = V(\bar{\phi}) + c$, where
$c$ is some constant. Consequently, nothing in this theory
depends on the value of the effective potentials $V(\phi)$
and $V(\bar{\phi})$ in their absolute minima $\phi_0$ and
$\bar{\phi}_0$. (Note, that $\phi_0=\bar{\phi}_0$ and $V(\phi_0) =
V(\bar{\phi}_0)$ due to the antipodal symmetry.) This is the
basic reason why it proves possible to solve the
cosmological constant problem in our model.

However, our main reason to invoke this new symmetry was not just to solve the
cosmological constant
problem. Just as the theory of mirror particles originally was proposed in
order to make the theory
CP-symmetric while maintaining CP-asymmetry in its observable sector, the
theory (\ref{eq43}) is
proposed in order to make the theory symmetric with respect
to the choice of the sign of energy. This removes the
old prejudice that, even though the overall change of sign of
the Lagrangian (i.e. both of its kinetic and potential
terms) does not change the solutions of the theory, one {\it
must say} that the energy of all particles is positive. This
prejudice was so strong, that many years ago physicists
preferred to quantize {\it particles} with {\it
negative energy} as {\it antiparticles} with {\it
positive energy}, which caused the appearance of such
meaningless concepts as negative probability. We wish to
emphasize that there is no problem to perform a consistent
quantization of theories which describe particles with
negative energy. All difficulties appear only when there
exist interacting species with both signs of energy. In our case no such
problem exists, just as there is no problem of antipodes
falling down from the opposite side of the earth. The reason is that
the fields $\bar{\phi}(y)$ do not interact with the fields
$\phi(x)$, and the equations of motion for the fields
$\bar{\phi}(y)$ are the same as for the fields $\phi(x)$ (the
overall minus sign in front of $L(\bar{\phi}(y))$ does not
change the Lagrange equations). Similarly, gravitons from
different universes do not interact with each other.
However, some interaction between the two universes does
exist. Indeed, the Einstein equations in our case are:
\begin{eqnarray}\label{eq44}
R_{\mu\nu} (x)-\frac{1}{2}g_{\mu\nu}R(x)
 = - 8\pi GT_{\mu\nu}(x) - g_{\mu\nu} \langle \frac{1}{2}
R(y) + 8\pi GL(\bar{\phi}(y))\rangle,
\end{eqnarray}
\begin{eqnarray}\label{eq45}
R_{\alpha\beta}(y) -\frac{1}{2}\bar{g}_{\alpha\beta} R (y)
 =-8\pi GT_{\alpha\beta} (y)-\bar{g}_{\alpha\beta}
\langle\frac{1}{2} R(x)+ 8\pi GL(\phi(x))\rangle
\end{eqnarray}
Here $T_{\mu\nu} $ is the energy-momentum tensor of the
fields $\phi(x), T_{a\beta}$ is the energy-momentum tensor
of the fields $\bar{\phi}(y)$, the sign of averaging means
\begin{equation}\label{eq46}
\langle R(x)\rangle =\frac{\int d4 x\sqrt{g(x)} R(x)}
{\int d^4x\sqrt{g(x)}},
\end{equation}

\begin{equation}\label{eq47}
\langle R(y)\rangle = \frac{\int d^4y\sqrt{\bar{g}(y)} R(y)}
{\int d^4y\sqrt{\bar{g}(y)}},
\end{equation}
and similarly for $\langle L(x)\rangle$ and $\langle
L(y)\rangle$. Thus, the novel feature of the theory (\ref{eq43}) is
the existence of a {\it global} interaction between the
universes X and Y: The integral {\it over the whole history}
of the Y-universe changes the vacuum energy density of the
X-universe.

In general, the computation of the averages of the type
(\ref{eq46}), (\ref{eq47}) may be a rather sophisticated problem.
Fortunately, however, in the inflationary theory
(at least, if the universe is not self-reproducing, see
below), this task can be rather trivial. Namely, the universe
after inflation becomes almost flat and its lifetime becomes
exponentially large.
In such a case, the dominant contribution to the
average values $\langle R\rangle$ and $\langle L \rangle$
comes from the late stages of the universe evolution at
which the fields $\phi (x)$ and $\phi(\bar{a})$ relax near the
absolute minima of their effective potentials. As a result,
the average value of $-L(\phi(x))$ almost exactly
 coincides with the
value of the effective potential $V(\phi)$ in its absolute
minimum at $\phi=\phi_0$, and the averaged value of the
curvature scalar $R(x)$ coincides with its value at the late
stages of the universe evolution, when the universe
transforms to the state corresponding to the absolute
minimum of $V(\phi)$. Similar results are valid for the
average values of $-L(\bar{\phi}(y))$ and of $R(y)$ as well.
In such a case one can easily show (Linde, 1988) that at the late
stages of the universe evolution, when the fields $\phi(x)$
and $\bar{\phi}(y)$ relax near the absolute minima of their
effective potentials, the {\em effective} cosmological
constant automatically vanishes,
\begin{equation}\label{eq48}
R(x) = - R(y) = \frac{32}{3}\pi G[V(\phi_0)-V(\bar{\phi}_0)]=0
\end{equation}

This model provided the first example of a theory with a non-local interaction of universes. It inspired the baby-universe scenario, and it was forgotten when the baby-universe scenario failed. However, this model is based on a completely different principle, so it should be considered quite independently.

There are several problems with this model that should be addressed before taking it too seriously. First of all, in order to solve the cosmological constant problem in our universe we added a new universe with negative energy density. At the first glance, this may not seem very economical. However, during the last  several years the idea that we may have several different interacting universes became very popular in the context of the brane world scenario (Arkani-Hamed {\it et al}, 1998,2000; Antoniadis {\it et al}, 1999; Randall and Sundrum, 1999).  The cancellation of the effective cosmological constant on our brane (our universe) is often achieved by the introduction of the negative tension brane (the universe with a negative energy density), see e.g. (Randall and Sundrum, 1999).  It is not quite clear whether any symmetry can protect this cancellation against radiative corrections in the brane world scenario. Meanwhile in our case the theory is fully symmetric with respect to the choice of the sign of energy, which may protect the cosmological constant against radiative corrections.

The second problem is more complicated. 
If the universe is self-reproducing, one may encounter
difficulties when computing the averages (\ref{eq46}), (\ref{eq47}),
since they may become dominated by eternally inflating parts of the universe with large $V(\phi)$.  One can avoid this complication in inflationary theories where
$V(\phi)$ grows rapidly enough at large $\phi$, since
there will be no universe self-reproduction in such
theories. 

Finally, the cosmological observations indicate that the universe is accelerating as if it has a miniscule positive vacuum energy $V(\phi) \sim 10^{-123} M_p^4$. Thus we need to make the vacuum energy cancellation non-exact. This is quite possible: as we said, the average value of $-L(\phi(x))$ {\it almost} exactly
 coincides with the
value of the effective potential $V(\phi)$ in its absolute
minimum at $\phi=\phi_0$. Also, if $V(\phi)$ is very flat near its minimum, like in the usual dark energy models, we may move to the minimum very slowly and at any given moment we will still have a small non-compensated positive vacuum energy.

We do not know whether this simple model is going to survive in the future. But this example shows that the multiverse scenario may provide us with 
new unexpected possibilities that should be considered very seriously. 

Now we will make a step back and discuss the anthropic approach to the cosmological constant problem.

\section{Cosmological constant, dark energy, and the anthropic principle}

The first attempt to solve the cosmological constant problem using the anthropic principle in the context of inflationary cosmology was made in (Linde, 1984b,1986b).  
The simplest way to do it is to consider inflation driven by the scalar field $\phi$ (the inflaton field)  and mimic the cosmological constant by the very flat potential of the second scalar field, $\Phi$. The simplest potential of this type is the linear potential (Linde, 1986b)
\begin{equation}\label{quint}  
V(\Phi) =\alpha M_p^3 \Phi \ .
\end{equation} 
If $\alpha$ is sufficiently small, $\alpha < 10^{-122}$, the potential 
$V(\Phi)$ is so flat that the field $\Phi$ practically does not change during the last $10^{10}$ years, its kinetic energy is very small, so at the present stage of the evolution of the universe 
its total potential energy $V(\Phi)$ acts exactly as a cosmological constant. This model was one of the first examples of what later became known as quintessence, or dark energy.

Even though the energy density of the field $\Phi$ practically does not change at the present time, it changed substantially during inflation. Since $\Phi$ is a massless field, it experienced quantum jumps with the amplitude $H/2\pi$ during each time $H^{-1}$. These jumps move the field $\Phi$ in all possible directions. In the context of the eternal inflation scenario this implies that the field becomes randomized by quantum fluctuations: The universe becomes divided into infinitely large number of exponentially large parts containing all possible values of the field $\Phi$. In other words, the universe becomes divided into infinitely large number of `universes' with all possible values of the effective cosmological constant $\Lambda =V(\Phi)+V(\phi_0)$, where $V(\phi_0)$ is the energy density of the inflaton field $\phi$ in the minimum of its effective potential. This quantity may change from $-M_p^4$ to $+M_p^4$ in different parts of the universe, but we can live only in the `universes' with $|\Lambda| \la O(10)\rho _0 \sim 10^{-28}$ g/cm$^3$, where $\rho_0$ is the present energy density in our part of the universe.

Indeed, if $\Lambda \la -10^{-28}$ g/cm$^3$, the universe collapses within the time much smaller than the present age of the universe $\sim 10^{10}$ years (Linde, 1984b,1986b; Barrow and Tipler, 1986). On the other hand, if $\Lambda \gg 10^{-28}$ g/cm$^3$, the universe at present would expand exponentially fast, energy density of matter would be exponentially small, and life as we know it would be impossible (Linde, 1984b,1986b). This means that we can live only in those parts of the universe where the cosmological constant does not differ too much from its presently observed value $|\Lambda| \sim \rho _0$.

This approach constituted the basis for many subsequent attempts to solve the cosmological constant problem using the anthropic principle in inflationary cosmology (Weinberg, 1987; Linde, 1990a; Vilenkin, 1995b; Martel {\it et al}, 1998; Garriga and Vilenkin, 2000,2001b,2002). 

At first glance, an introduction of the miniscule parameter $\alpha < 10^{-122}$ does not provide a real  explanation of the equally miniscule cosmological constant $|\Lambda| \sim \rho _0 \sim 10^{-123} M_p^4$. However, exponentially small parameters like that may easily appear due to nonperturbative effects. One could even think that a similar exponential suppression may be the true reason why $|\Lambda|$ is so small. But there are many large contributions to $\Lambda$, due to quantum gravity, due to spontaneous symmetry breaking in GUTs and in the electroweak theory, due to supersymmetry breaking, QCD effects, etc. One could appeal to the nonperturbative exponential smallness of $\Lambda$ only if all large contributions to the vacuum energy miraculously cancel, like in the model considered in the previous section. And even if this cancellation is achieved, we still need to explain why $|\Lambda|$ is suppressed exactly to the level when it becomes of the same order as   the present energy density of the universe. This coincidence problem becomes resolved in the theory (\ref{quint}) for all sufficiently small $\alpha$; instead of the fine-tuning of $\alpha$ we simply need it to be sufficiently strongly suppressed. A very clear discussion of the issue of fine-tuning versus exponential suppression  can be found in (Garriga and Vilenkin, 2000)  in application to a similar model with the potential $\rho_\Lambda \pm m^2 \Phi^2/2$ with $m^2 \ll 10^{-240} M_p^6 |\rho_\Lambda|^{-1}$.

Alternative approaches based on the anthropic principle are described in (Bousso and  Polchinski, 2001; Feng  {\it et al}, 2001; Banks {\it et al}, 2001).
One can also use a more general approach outlined in Section \ref{multi} and consider a baby-universe scenario, or a multiverse consisting of different inflationary universes with different values of the cosmological constant in each of them (Linde, 1989,1990a,1991). In this case one does not need to consider extremely flat potentials, but the procedure of comparing probabilities to live in different universes with different $\Lambda$ becomes more ambiguous (Vilenkin, 1995;  Garcia-Bellido and Linde, 1995).  However, if one makes the simplest  assumption that the universes with different values of $\Lambda$ are equally probable, one obtains an anthropic solution of the cosmological constant problem without any need of introducing extremely small parameters $\alpha < 10^{-122}$ or $m^2 \ll 10^{-240} M_p^6 |\rho_\Lambda|^{-1}$.

  The constraint $\Lambda \ga -10^{-28}$ g/cm$^3$ still remains the strongest constraint on the negative cosmological constant; for the recent developments related to this constraint see (Kallosh and Linde, 2002; Garriga and Vilenkin, 2002). Meanwhile, the constraint on the positive cosmological constant, $\Lambda \la 10^{-28}$ g/cm$^3$, was made much more precise and accurate in the subsequent works.

In particular, Weinberg  pointed out that the process of galaxy formation occurs only up to the moment when the cosmological constant begins to dominate the energy density of the universe and the universe enters the stage of late-time inflation (Weinberg, 1987). For example, one may consider galaxies  formed  at $z \ga
4$, when the energy density of the universe was 2  orders of
magnitude greater than it is now. Such galaxies would not form if  $\Lambda
\ga 10^2 \rho_0 \sim 10^{-27}$ g/cm$^3$.

The next important step  was made  in a series of works (Efstathiou, 1995; Vilenkin, 1995b; Martel {\it et al}, 1998; Garriga and Vilenkin, 2000,2001b,2002; Bludman and Roos, 2002). The
authors considered not only our own galaxy, but all other galaxies that
could harbor  life of our type. This would include not only the existing galaxies but also the
galaxies that are being formed at the present epoch. Since the energy density at later stages of the evolution of the universe becomes smaller, even a very small cosmological constant may disrupt the late-time galaxy formation, or may prevent the growth of existing galaxies.   This allows to strengthen the constraint on the cosmological constant. According to (Martel {\it et al}, 1998), the probability that an astronomer in any of the universes would find the presently observed ratio $\Lambda/\rho_0$ as small as $0.7$ ranges from $5\%$ to $12\%$, depending on various assumptions. For some models based on extended supergravity, the anthropic constraints can be strengthened even further (Kallosh and Linde, 2002).

\section{Problem of calculating the probabilities}

As we see, the anthropic principle can be extremely useful in resolving some of the most profound problems of modern physics. However, to make this principle  more  quantitative, one
should find a proper way to calculate the probability to live in a universe of a given type. This step is not quite trivial. One may consider the probability of  quantum creation of the universe `from nothing' (Hartle and Hawking, 1983; Linde, 1984a; Vilenkin, 1984), or the results of the `baby universe' theory (Coleman, 1988a,1988b), or the results based on  the theory of the self-reproduction of the universe and quantum cosmology (Linde {\it et al}, 1994; Garcia-Bellido {\it et al}, 1994; Vilenkin, 1995; Vanchurin {\it et al}, 2000; Garcia-Bellido and Linde, 1995; Linde and Mezhlumian, 1996; Garriga and Vilenkin, 2001). Unfortunately, these methods   are based on different assumptions, and the results of some of these works significantly differ from each other. This may be just a temporary setback. For example, in our opinion, an interpretation of Euclidean quantum gravity used in  (Hartle and Hawking, 1983, Coleman, 1988a,1988b) is not quite convincing. The method proposed in (Turok, 2002) is basically equivalent to the investigation of the probability distribution in comoving coordinates $P_c(\phi,t)$ (Linde, 1990a). This approach ignores information about most of the observers living in our universe, so it  can hardly have any relation to the standard anthropic considerations and misses the effect of the self-reproduction of the universe. An investigation of creation of the universe `from nothing' (Linde, 1984a; Vilenkin, 1984) can be very useful, but I believe that it should be considered only as a part of the more general approach based on the stochastic approach to inflation.

It is more difficult to make a definite choice between the different answers provided by the different methods of interpretation of the results obtained by the stochastic approach to inflation (Starobinsky, 1986; Linde {\it et al}, 1994; Garcia-Bellido {\it et al}, 1994; Vilenkin, 1995;  Garcia-Bellido and Linde, 1995; Linde and Mezhlumian, 1996; Vanchurin {\it et al}, 2000; Garriga and Vilenkin, 2001a). We believe that all of these different answers  in a certain sense are correct; it is the choice of the questions that remains problematic.

To explain our point of view, let us study an example related to demographics. One may want to know what is the average age of a person living now on the Earth. In order to find it, one should take the sum of the ages of all people and divide it by their total number. Naively, one could expect that the
result of the calculation should be equal to $1/2$ of the life expectancy. However, the actual result will be much smaller. Because of the exponential growth of the population, the main contribution to the
average age will be given by very young people. Both answers (the average age of a person, and a half of the life expectancy) are correct despite the fact that they are different. None of these answers is any better; they are different because they  address different questions.  Economists may want to know the average age in order to make their projections. Meanwhile each of us, as well as the people from  the insurance industry, may be more interested in the life expectancy.

Similarly, the calculations performed in (Linde {\it et al}, 1994; Garcia-Bellido {\it et al}, 1994; Vilenkin, 1995;  Garcia-Bellido and Linde, 1995; Linde and Mezhlumian, 1996; Vanchurin {\it et al}, 2000; Garriga and Vilenkin, 2001a) dissect all
possible outcomes of the evolution of the universe (or the multiverse) in many different ways. (Unlike the method suggested in (Turok, 2002), these methods cover the whole universe rather that its infinitesimally small part.) Each of these ways is quite legitimate and leads to correct results, but some additional input is required in order to understand which of these results, if any, is most closely related to the anthropic principle. 

In the meantime one may take a pragmatic point of view and consider this investigation as a kind
of `theoretical experiment.'  We may try to use probabilistic
considerations in a trial-and-error approach. If we get unreasonable
results, this may serve as an indication that we are using quantum
cosmology incorrectly. However, if some particular proposal for the
probability measure will allow us to solve
certain problems which could not be solved in any other way, then we
will have a reason to believe that  we are moving in the right direction. But we are not sure that any real progress in this direction can be reached and we will be able to learn how to calculate the probability to live  in one of the many universes without having a good idea of what is life and what is consciousness (Linde, 1990a;  Garcia-Bellido and Linde, 1995; Linde and Mezhlumian, 1996; Linde {\it et al}, 1996).

A healthy scientific conservatism usually forces us to disregard all metaphysical subjects that  seem unrelated to our research. However, in order to make sure that this conservatism is really healthy, from time to time one should take a risk to abandon some of the standard assumptions. This may allow us  either to reaffirm our previous position, or to find some possible limitations of our earlier point of view.

\section{Does consciousness matter?}

 A good starting point for our brief discussion of consciousness is quantum
cosmology, the theory that tries to unify cosmology and quantum
mechanics.

If quantum mechanics is universally correct, then one may try to apply it to the universe in order to find its wave function. This would allow us find out which
events are probable and which are not. However, it often leads to
paradoxes. For example,  the essence
of the Wheeler-DeWitt equation (DeWitt, 1967), which is the Schr\"{o}dinger
equation for the wave function of the universe, is that this wave
function {\it does not depend on time}, since the total
Hamiltonian of the universe, including the Hamiltonian of the
gravitational field, vanishes identically. This result was
obtained in 1967 by   Bryce
DeWitt. Therefore if one would wish to describe the evolution of
the universe with the help of its wave function, one would be in
trouble: {\it The universe as a whole does not change in time}.

The resolution of this paradox  suggested by   Bryce
DeWitt is rather instructive (DeWitt, 1967). The notion
of evolution is not applicable to the universe as a whole since
there is no external observer with respect to the universe, and
there is no external clock  that does not belong to the
universe. However, we do not actually ask why the universe {\it
{as a whole}} is evolving. We are just trying
to understand our own experimental data. Thus, a more precisely
formulated question is {\it why do   we see } the universe
evolving in time in a given way. In order to answer this question
one should first divide the universe into two main pieces: ~i) an
observer with his clock and other measuring devices and ~ii) the
rest of the universe. Then it can be shown that the wave function
of the rest of the universe does depend on the state of the clock
of the observer, i.e. on his `time'. This time dependence in
some sense is `objective':   the results obtained
by different (macroscopic) observers living in the same quantum
state of the universe and using sufficiently good (macroscopic)
measuring apparatus agree with each other.

Thus we see that without
introducing an observer, we have a dead universe, which does not
evolve in time.   
This example demonstrates an unusually important role played by
the concept of an observer in quantum cosmology.  John Wheeler underscored the complexity of the
situation, replacing the word {\it observer} by the word {\it
participant}, and introducing such terms as a `self-observing
universe'.   

Most of the
time, when discussing quantum cosmology, one can remain entirely
within the bounds set by purely physical categories, regarding an
observer simply as an automaton, and not dealing with questions of
whether he/she/it has consciousness or feels anything during the process
of observation. This limitation is harmless for many practical
purposes. But we cannot rule out the possibility  
that carefully avoiding the concept of consciousness in quantum
cosmology may lead to an artificial narrowing of our outlook.  

Let us remember an example from the history of science that may
be rather instructive in this respect. Prior to the
invention of  the general theory of relativity, space, time, and
matter seemed to be three fundamentally different entities.  Space was thought to be a kind of three-dimensional
coordinate grid which, when supplemented by clocks, could be used
to describe the motion of matter.  Space-time   possessed no
intrinsic degrees of freedom,  it  played
secondary role as a tool for the description of the
truly substantial material world.

The general theory of relativity brought with it a decisive change
in this point of view.  Space-time and matter were found to be
interdependent, and there was no longer any question which one of the two is
more fundamental.  Space-time was also found to
have its own inherent degrees of freedom, associated with
perturbations of the metric --  gravitational waves.  Thus, space
can exist and change with time in the absence of electrons,
protons, photons, etc.;  in other words, in the absence of
anything that had previously (i.e., prior to general
relativity) been called  matter. Of course, one can simply extend the notion of matter, because, after all, gravitons (the quanta of the gravitational field) are real particles living in our universe. On the other hand,  the introduction of the gravitons provides us, at best, with a tool for an approximate (perturbative) description of the fluctuating geometry of space-time. This is completely opposite to the previous idea that space-time is only a tool for the description of matter.

A more recent trend, finally, has been toward a unified geometric
theory of all fundamental interactions, including gravitation. 
Prior to the end of the 1970's, such a program seemed
unrealizable;  rigorous theorems were proven on the impossibility
of unifying spatial symmetries with the internal symmetries of
elementary particle theory.  Fortunately, these theorems were
sidestepped after the discovery of supersymmetry and supergravity.  In
these theories, matter fields and space-time became unified within the 
general concept of superspace.

Now let us turn to consciousness. The standard assumption is that
 consciousness, just like space-time before the
invention of general relativity, plays a secondary, subservient
role, being  just a function of matter and a tool for
the description of the truly existing material world.  But let us
remember that our knowledge of the world begins not with matter
but with perceptions. I know for sure that my pain exists, my
`green' exists, and my `sweet' exists. I do not need any proof
of their existence, because these events are a part of me;
everything else is a theory. Later we find out that our
perceptions obey some laws, which can be most conveniently
formulated if we assume that there is some underlying reality
beyond our perceptions. This model of material world obeying laws
of physics is so successful that soon we forget about our starting
point and say that matter is the only reality, and perceptions are
nothing but a useful tool for the description of matter. This assumption is almost as natural (and maybe as false) as our previous assumption that space
is only a mathematical tool for the description of matter. 
We are substituting {\it reality} of our feelings by the
successfully working {\it theory} of  an independently existing
material world. And the theory is so successful that we almost
never think about its possible limitations.

Guided by the analogy with the gradual change of the concept of space-time, we would like to take a certain risk and  formulate several questions to which we do not yet have the answers (Linde, 1990a; Page, 2002):

Is it possible that consciousness, like space-time, has its
own intrinsic degrees of freedom, and that neglecting these will
lead to a description of the universe that is fundamentally
incomplete?  What if our perceptions are as real (or maybe, in a
certain sense, are even more real) than material objects? What if
my red, my blue, my pain, are really existing objects, not merely
reflections of the really existing material world? Is it possible
to introduce a `space of elements of consciousness,' and
investigate a possibility that consciousness may exist by itself,
even in the absence of matter, just like gravitational waves,
excitations of space, may exist in the absence of protons and
electrons?  

Note, that the gravitational waves usually are so small and interact with matter so weakly that we did not find any of them as yet. However, their existence is absolutely crucial for the consistency of our theory, as well as for our understanding of certain astronomical data. Could it be that consciousness is an equally important  part of the consistent picture of our world, despite the fact that so far  one  could safely ignore it in the description of the well studied physical processes?  Will it not turn out, with the further development of
science, that the study of the universe and the study of
consciousness are inseparably linked, and that ultimate
progress in the one will be impossible without progress in the
other?

Instead of discussing these issues here any further, we will return back to a more solid ground and concentrate on the consequences of eternal inflation and the  multiverse theory that do not depend on the details of their interpretation. As an example, we will discuss here two questions that for a long time were considered too complicated and metaphysical. We will see that the concept of the multiverse will allow us to propose possible answers to these questions.

\section{Why is mathematics so efficient?}

There is an old problem that bothered many people thinking about the foundations of mathematics: Why is mathematics so efficient in helping us to describe  our world and predict its evolution?

This question arises at the moment when one introduces numbers and uses them to count. Then a similar question appears when one introduces calculus and uses it to describe the motion of the planets. Somehow there are some rules that help us to operate with mathematical symbols and relate the results of these operations to the results of our observations. Why does it work so well?

Of course, one could always respond that it is just so. But let us consider several other questions of a similar type. Why is our universe so large? Why parallel lines do not intersect? Why different parts of the universe look so similar? Thirty years ago such questions would look too metaphysical to be considered seriously. Now we know that inflationary cosmology provides a possible answer to all of these questions. Let us try it again.

Before we do it, we should give at least one example of a universe where mathematics would be inefficient. Here it is. Suppose the universe can be in a stable or metastable vacuum state with a  Planckian density $\rho \sim M_p^4 \sim 10^{94}$ g/cm$^{3}$. According to quantum gravity, quantum fluctuations of space-time curvature in this regime are of the same order as the curvature itself. In simple terms, this means that the rulers are bending, shrinking and extending in a chaotic and unpredictable way due to quantum fluctuations, and this happens faster than one can measure the distance. The clocks are destroyed faster than one can measure the time. All records about the previous events become erased, so one cannot remember anything, record it, make a prediction and compare the prediction with experimental results.

A similar situations occurs in a typical non-inflationary closed universe. There is only one natural parameter of dimension of length in quantum gravity, $l_p = M_p^{-1}$, and only one natural parameter of dimension of energy density, $\rho_p = M_p^4$. If one considers a typical closed universe of a typical initial size $l_p$ with a typical initial density $\rho_p$, one can show that its total lifetime until it collapses  is $t \sim t_p = M_p^{-1} \sim 10^{-43}$ seconds, and throughout all of its short history the energy density remains of the order of $M_p^4$ or greater. Such a universe can incorporate just a few elementary particles (Linde, 1990a), so one cannot live there, cannot  build any measuring devices, record any events and use mathematics to describe events in such a universe. 

In the cases described above, mathematics would be rather inefficient  because it would not help anybody to relate different things and processes to each other. More generally, if the laws of physics inside some parts of the universe disallow formation of stable long-living structures, then mathematics will not be very useful there, and  there will be no observers (long-living conscious beings capable of remembering and thinking) who would be able to tell us about it.

Fortunately, among all possible  domains of the universe (or among all possible universes) there are some domains where inflation is possible. Energy density inside such  domain gradually drops down many orders of magnitude below $M_p^4$. These domains become exponentially large and can live for an exponentially long time.  Our life is possible only in those exponentially large domains (or universes) where  the laws of physics allow formation of stable long-living structures. The very concept of stability implies existence of  mathematical relations that can be used for the long-term predictions. The rapid development of the human race became possible only because we live in the universe where the long-term predictions are so useful and efficient that they allow us to survive in the hostile environment and win in the competition with other species.

To summarize, in the context of the multiverse theory, one can consider all possible universes with all possible laws of physics and mathematics.  Among all possible universes, we can live only in those where mathematics is efficient.

\section{Why quantum?}

Now we will discuss the famous Wheeler's question:  Why quantum?

 Before doing so, I would like to remember the question often asked by Zeldovich: Do we have any experimental evidence of  proton instability and baryon non-conservation?

In accordance to the unified theories of weak, strong and electromagnetic interactions,  protons and other baryons can be unstable. They can decay to  leptons. But the decay rate is so small that we still did not find any direct evidence of the proton instability. People were watching protons in thousands of tons of water, and did not find any of  them decaying. Thus the simple-minded answer to Zeldovich's question would be ``No.''

However, the true answer is different. To make it sound a little bit more challenging, I will formulate it in a way slightly different from the formulation used by Zeldovich, but conveying the same basic idea.

{\it The main experimental evidence of the baryon number non-conservation is provided by the fact that parallel lines do not intersect.}

What? Is it a joke? What is the relation?

Well, the fact that the parallel lines do not intersect and remain parallel to each other is a consequence of the spatial flatness of the universe. In a closed universe the parallel line would intersect, in an open universe they would diverge at infinity. The only known explanation of the flatness of the universe is provided by inflationary cosmology. This theory implies that at the  end of the exponential expansion of the universe,  the number density of all elementary particles becomes vanishingly small.  

All matter surrounding us was produced due to the decay of the scalar field after inflation (Dolgov and Linde, 1982; Abbott {\it et al}, 1982; Kofman {\it et al}, 1994,1997; Felder  {\it et al}, 2001). The density of protons in our part of the universe is much greater than the density of antiprotons. This means that at the present time  the total baryon number density  is not zero. It would be impossible to produce these baryons from the post-inflationary state with the vanishing baryon density if the baryon number were conserved.

Thus, the only available explanation of the observed flatness and homogeneity of the universe requires baryon number non-conservation. In this sense, the fact that the parallel lines do not intersect is an observational evidence of the proton instability.

This is a strange and paradoxical logic, but we must get used to it if we want to understand the properties of our universe.

Now let us return to Wheeler's question. At the first glance, this question   is so deep and metaphysical that we are not going to know the answer any time soon. However, in my opinion, the answer is pretty simple.

The only known way to explain why our universe is so large,  flat, homogeneous and isotropic requires inflation. As we just said, after inflation the universe becomes empty.  All matter in the universe was produced due to {\it quantum} processes after the end of inflation. All galaxies were produced by {\it quantum fluctuations} generated at the last stages of inflation. There would be no galaxies and no matter in our universe if not for the {\it quantum effects}. One can formulate this result in the following way:

{\it Without inflation, our universe would be ugly. Without quantum, our universe would be empty.} 

But there is something else here. As we already discussed in Section \ref{eter}, {\it quantum} fluctuations lead to the eternal process of self-reproduction of the inflationary universe. 

{\it  Quantum effects combined with  inflation  make the universe infinitely large and immortal.}

This provides a possible answer to Wheeler's question.

\
 
\
 
 Isn't it amazing that different, apparently unrelated things can match together to form a beautiful and self-consistent pattern? Are we uncovering the universal truth or simply allow this beauty to deceive us? This is one of the questions that will remain with us for some time. We need to move  carefully and slowly, constantly keeping in touch with solid and well established facts, but from time to time allowing ourselves to satisfy our urge to speculate, following the steps of John Wheeler.

\newpage

\section*{
References}


~~~~Abbott, L.~F., Farhi, E. and Wise, M.~B.  (1982)
``Particle Production In The New Inflationary Cosmology,''
Phys.\ Lett.\ B {\bf 117}, 29.

Albrecht, A. and Steinhardt, P.~J. (1982)
``Cosmology For Grand Unified Theories With Radiatively Induced Symmetry Breaking,''
Phys.\ Rev.\ Lett.\  {\bf 48}, 1220.

Antoniadis, I., Arkani-Hamed, N., Dimopoulos, S. and Dvali, G.~R. (1998)
``New dimensions at a millimeter to a Fermi and superstrings at a TeV,''
Phys.\ Lett.\ B {\bf 436}, 257  
[arXiv:hep-ph/9804398].

Arkani-Hamed, N., Dimopoulos, S. and Dvali, G.~R. (1998)
``The hierarchy problem and new dimensions at a millimeter,''
Phys.\ Lett.\ B {\bf 429}, 263
[arXiv:hep-ph/9803315]. 

Arkani-Hamed, N., Dimopoulos, S.,  Dvali, G.~R. and  Kaloper, N.  (2000)
``Manyfold universe,''
JHEP {\bf 0012}, 010 
[arXiv:hep-ph/9911386].

Banks, T. (1988)
``Prolegomena To A Theory Of Bifurcating Universes,''
Nucl.\ Phys.\ B {\bf 309}, 493.

Banks, T.,   Dine, M.,  and  Motl, L. (2001)
``On anthropic solutions of the cosmological constant problem,''
JHEP {\bf 0101}, 031
[arXiv:hep-th/0007206].

Bardeen, J. M., Steinhardt, P.~J. and Turner, M.~S. (1983) ``Spontaneous Creation Of Almost Scale - Free Density Perturbations In An Inflationary Universe,''
Phys.\ Rev.\ D {\bf 28}, 679.

Barrow,J. D. and Tipler, F. J. (1986) {\it The Anthropic Cosmological Principle}, Oxford University Press, New York.

Bludman, S.~A. and Roos, M.  (2002)
``Quintessence cosmology and the cosmic coincidence,''
Phys.\ Rev.\ D {\bf 65}, 043503
[arXiv:astro-ph/0109551].

Borde, A., Guth, A.~H. and Vilenkin, A. (2001)
``Inflation is not past-eternal,''
arXiv:gr-qc/0110012.

Bousso, R. and  Polchinski, J.  (2000)
 ``Quantization of four-form fluxes and dynamical neutralization of the  cosmological constant,''
JHEP {\bf 0006}, 006
[arXiv:hep-th/0004134].

Coleman, S.~R. (1988a)
``Black Holes As Red Herrings: Topological Fluctuations And The Loss Of Quantum Coherence,''
Nucl.\ Phys.\ B {\bf 307}, 867.

Coleman, S.~R. (1988b)
``Why There Is Nothing Rather Than Something: A Theory Of The Cosmological Constant,''
Nucl.\ Phys.\ B {\bf 310}, 643.

DeWitt, B.~S. (1967)
``Quantum Theory Of Gravity. 1. The Canonical Theory,''
Phys.\ Rev.\  {\bf 160}, 1113 (1967).

Dolgov, A.~D. and Linde, A. D.  (1982)
``Baryon Asymmetry In Inflationary Universe,''
Phys.\ Lett.\ B {\bf 116}, 329.

Donoghue, J.~F. (2000)
 ``Random values of the cosmological constant,''
JHEP {\bf 0008}, 022
[arXiv:hep-ph/0006088].

Efstathiou, G. (1995) M.N.R.A.S. {\bf 274},
L73.

Felder, G., Garcia-Bellido, J., P.~B.~Greene, Kofman, L.,   
Linde, A. D. and Tkachev, I.  (2001)
 ``Dynamics of symmetry breaking and tachyonic preheating,'' Phys. Rev. Lett. {\bf 87}, 011601,   
hep-ph/0012142.

Feng, J.~L., March-Russell, J.,  Sethi, S. and  Wilczek, F.  (2001)
``Saltatory relaxation of the cosmological constant,''
Nucl.\ Phys.\ B {\bf 602}, 307
[arXiv:hep-th/0005276]. 

Garcia-Bellido, J., Linde, A. D. and Linde, D. A.  (1994)
``Fluctuations of the gravitational constant in the inflationary Brans-Dicke cosmology,''
Phys.\ Rev.\ D {\bf 50}, 730
[arXiv:astro-ph/9312039].

Garcia-Bellido, J. and Linde, A. D. (1995)
``Stationarity of inflation and predictions of quantum cosmology,''
Phys.\ Rev.\ D {\bf 51}, 429
[arXiv:hep-th/9408023].

Garriga, J. and Vilenkin, A.  (2000),
``On likely values of the cosmological constant,''
Phys.\ Rev.\ D {\bf 61}, 083502
[arXiv:astro-ph/9908115].

Garriga, J. and Vilenkin, A.  (2001a),
 ``A prescription for probabilities in eternal inflation,''
Phys.\ Rev.\ D {\bf 64}, 023507
[arXiv:gr-qc/0102090].

Garriga, J. and Vilenkin, A. (2001b),
``Solutions to the cosmological constant problems,''
Phys.\ Rev.\ D {\bf 64}, 023517 
[arXiv:hep-th/0011262].

Garriga, J. and Vilenkin, A. (2002)
``Testable anthropic predictions for dark energy,''
arXiv:astro-ph/0210358.

Giddings, S.~B. and Strominger, A. (1988)
``Loss Of Incoherence And Determination Of Coupling Constants In Quantum Gravity,''
Nucl.\ Phys.\ B {\bf 307}, 854.

Giddings, S.~B. and Strominger, A. (1989)
``Baby Universes, Third Quantization And The Cosmological Constant,''
Nucl.\ Phys.\ B {\bf 321}, 481.

Guth, A. H. (1981)
``The Inflationary Universe: A Possible Solution To The Horizon And Flatness Problems,''
Phys.\ Rev.\ D {\bf 23}, 347.

Guth, A.~H. and S.~Y.~Pi (1982)
``Fluctuations In The New Inflationary Universe,''
Phys.\ Rev.\ Lett.\  {\bf 49}, 1110.

Hawking, S.~W.  (1982)
``The Development Of Irregularities In A Single Bubble Inflationary Universe,''
Phys.\ Lett.\ B {\bf 115}, 295.

Hawking, S.~W. (1984)
``The Cosmological Constant Is Probably Zero,''
Phys.\ Lett.\ B {\bf 134}, 403.

Hartle, J.~B. and Hawking, S.~W. (1983).
``Wave Function Of The Universe,''
Phys.\ Rev.\ D {\bf 28}, 2960

Kallosh, R. and Linde, A. D. (2002)
``M-theory, cosmological constant and anthropic principle,''
arXiv:hep-th/0208157.

Kofman, L., Linde, A. D. and Starobinsky, A.~A. (1994)
``Reheating after inflation,''
Phys.\ Rev.\ Lett.\  {\bf 73}, 3195
[arXiv:hep-th/9405187].

Kofman, L., Linde, A. D. and Starobinsky, A.~A. (1997)
``Towards the theory of reheating after inflation,''
Phys.\ Rev.\ D {\bf 56}, 3258
[arXiv:hep-ph/9704452].

Kolb, E. W.  and Turner, M. S. (1990) {\it The Early Universe}, Addison-Wesley,
New
York.

Linde, A. D. (1982a)
``A New Inflationary Universe Scenario: A Possible Solution Of The Horizon, Flatness, Homogeneity, Isotropy And Primordial Monopole Problems,''
Phys.\ Lett.\ B {\bf 108}, 389. 

Linde, A. D. (1982b)
``Nonsingular Regenerating Inflationary Universe,''
Print-82-0554 (Cambridge University).

Linde, A. D. (1982c)
``Scalar Field Fluctuations In Expanding Universe And The New Inflationary Universe Scenario,''
Phys.\ Lett.\ B {\bf 116}, 335 (1982).

Linde, A. D. (1983a)
``The New Inflationary Universe Scenario,''
In: {\it  The Very Early Universe}, ed. G.W. Gibbons, S.W. Hawking and S.Siklos,  pp. 205-249 
Cambridge University Press.

Linde, A. D. (1983b) ``Chaotic Inflation,'' Phys. Lett. {\bf 129B},  177.

Linde, A. D.  (1983c)
``Inflation Can Break Symmetry In Susy,''
Phys.\ Lett.\ B {\bf 131}, 330.

 Linde, A. D. (1984a)
``Quantum Creation Of The Inflationary Universe,''
Lett.\ Nuovo Cim.\  {\bf 39}, 401.

Linde, A. D. (1984b) ``The Inflationary Universe,'' Rep. Prog. Phys. {\bf 47},   925. 

Linde, A. D. (1985)
``The New Mechanism Of Baryogenesis And The Inflationary Universe,''
Phys.\ Lett.\ B {\bf 160}, 243.

Linde, A. D. (1986a)
``Eternally Existing Self-reproducing Chaotic Inflationary Universe,''
Phys.\ Lett.\ B {\bf 175}, 395.

Linde, A. D. (1986b)
``Inflation And Quantum Cosmology,''
Print-86-0888
(June 1986),
in: {\it  Three hundred years of gravitation}, (Eds.: Hawking, S.W.  and Israel, W., Cambridge Univ. Press, 1987), 604-630.

Linde, A. D. (1987a) ``Particle Physics and Inflationary Cosmology,'' Physics Today {\bf 40}, 61.

Linde, A. D. (1987b)
``Inflation And Axion Cosmology,''
Phys.\ Lett.\ B {\bf 201}, 437 (1988).

Linde, A. D. (1988)
``The Universe Multiplication And The Cosmological Constant Problem,''
Phys.\ Lett.\ B {\bf 200}, 272.

Linde, A. D. (1989)
``Life After Inflation And The Cosmological Constant Problem,''
Phys.\ Lett.\ B {\bf 227}, 352.

Linde, A. D. (1990a) {\it Particle Physics and Inflationary Cosmology}, Harwood 
Academic Publishers, Chur, Switzerland. 

Linde, A. D.  (1990b)
``Extended Chaotic Inflation And Spatial Variations Of The Gravitational Constant,''
Phys.\ Lett.\ B {\bf 238}, 160. 

Linde, A. D.  (1991)
``Inflation And Quantum Cosmology,''
Phys.\ Scripta {\bf T36}, 30.

Linde, A. D. (1998)
``Quantum creation of an open inflationary universe,''
Phys.\ Rev.\ D {\bf 58}, 083514
[arXiv:gr-qc/9802038].

Linde, A. D., Linde, D. A.  and Mezhlumian, A. (1994)
``From the Big Bang theory to the theory of a stationary universe,''
Phys.\ Rev.\ D {\bf 49}, 1783
[arXiv:gr-qc/9306035].

Linde, A. D., Linde, D. A. and Mezhlumian, A. (1996)
``Nonperturbative Amplifications of Inhomogeneities in a Self-Reproducing Universe,''
Phys.\ Rev.\ D {\bf 54}, 2504
[arXiv:gr-qc/9601005].

Linde, A. D. and Mezhlumian, A. (1996)
``On Regularization Scheme Dependence of Predictions in Inflationary Cosmology,''
Phys.\ Rev.\ D {\bf 53}, 4267
[arXiv:gr-qc/9511058].

Linde, A. D. and Zelnikov, M.~I. (1988)
``Inflationary Universe With Fluctuating Dimension,''
Phys.\ Lett.\ B {\bf 215}, 59.

Martel, H., Shapiro P.~R.~and Weinberg, S. (1998)
``Likely Values of the Cosmological Constant,'' Astrophys. J. {\bf 492}, 29.
arXiv:astro-ph/9701099.

Mukhanov, V.~F.  (1985)
``Gravitational Instability Of The Universe Filled With A Scalar Field,''
JETP Lett.\  {\bf 41}, 493.

Mukhanov, V.~F. and G.~V.~Chibisov,
``Quantum Fluctuation And `Nonsingular' Universe,''
JETP Lett.\  {\bf 33}, 532 (1981)
[Pisma Zh.\ Eksp.\ Teor.\ Fiz.\  {\bf 33}, 549 (1981)]. 

Page, D.~N.~ (2002)
``Mindless Sensationalism: A Quantum Framework for Consciousness,''  in: {\it Consciousness: New Philosophical Essays}, Eds. Q. Smith and A. Jokic. Oxford, Oxford University Press.
arXiv:quant-ph/0108039.

Randall, L, and  Sundrum, R.  (1999)
``A large mass hierarchy from a small extra dimension,''
Phys.\ Rev.\ Lett.\  {\bf 83}, 3370
[arXiv:hep-ph/9905221].

Rees, M. (2000) {\it  Just Six Numbers: The Deep Forces that Shape the Universe}, 
Basic Books, Perseus Group,
New York.

Rozental, I.L. (1988) {\it Big Bang, Big Bounce}, Springer Verlag.

Starobinsky, A.A. (1980)
``A New Type Of Isotropic Cosmological Models Without Singularity,''
Phys.\ Lett.\ B {\bf 91}, 99.

Starobinsky, A.~A.  (1982)
``Dynamics Of Phase Transition In The New Inflationary Universe Scenario And Generation Of Perturbations,''
Phys.\ Lett.\ B {\bf 117}, 175.

Starobinsky, A.~A.  (1986) ``Stochastic De Sitter (Inflationary) Stage In The Early Universe,'' in: {\it Current Topics in Field
Theory, Quantum Gravity and Strings}, Lecture Notes in Physics, eds.
H.J. de Vega and N. Sanchez (Springer, Heidelberg) {\bf 206},
p. 107.

Steinhardt, P. (1983)
``Natural Inflation,''
In: {\it  The Very Early Universe}, ed. G.W. Gibbons, S.W. Hawking and S.Siklos, 
Cambridge University Press.

Tegmark, M. (1998)
``Is *the theory of everything* merely the ultimate ensemble theory?''
Annals Phys.\  {\bf 270}, 1
[arXiv:gr-qc/9704009].

Turok, N. (2002)
``A Critical Review Of Inflation,''
Class.\ Quant.\ Grav.\  {\bf 19}, 3449 (2002).

Vanchurin, V., Vilenkin, A.  and Winitzki, S. (2000)
``Predictability crisis in inflationary cosmology and its resolution,''
Phys.\ Rev.\ D {\bf 61}, 083507
[arXiv:gr-qc/9905097].

Vilenkin, A.  (1983)
``The Birth Of Inflationary Universes,''
Phys.\ Rev.\ D {\bf 27}, 2848.

Vilenkin, A.  (1984)
``Quantum Creation Of Universes,''
Phys.\ Rev.\ D {\bf 30}, 509.

Vilenkin, A.  (1995)
``Predictions From Quantum Cosmology,''
Phys.\ Rev.\ Lett.\  {\bf 74}, 846
[arXiv:gr-qc/9406010].

Vilenkin, A. and Ford, L.~H. (1982)
``Gravitational Effects Upon Cosmological Phase Transitions,''
Phys.\ Rev.\ D {\bf 26}, 1231.

Weinberg, S. (1987) ``Anthropic Bound On The Cosmological Constant,'' Phys. Rev. Lett. {\bf 59}, 2607. 

Wheeler, J A. (1990) ``Information, Physics, Quantum: The Search for Links,'' in: {\it Complexity, Entropy and the Physics of Information},   ed. W.H Zurek, Addison-Wesley,
pp. 3-28.

\end{document}